\documentclass[preprint,amsmath,amssymb,aps,floatfix,showkeys,showpacs,
]{revtex4-1}

\usepackage[utf8]{inputenc}
\usepackage[T1]{fontenc}
\usepackage{amsmath,amssymb,color,graphicx,bm}
\usepackage{times}
\usepackage{dcolumn}
\usepackage{marginnote}

\newcommand{\bal}{\boldsymbol{\alpha}}
\newcommand{\bbe}{\boldsymbol{\beta}}
\newcommand{\bga}{\boldsymbol{\gamma}}
\newcommand{\bde}{\boldsymbol{\delta}}

\begin{document}

\title[Boussinesq's equations]{Boussinesq's equations for (2+1)-dimensional surface gravity waves in an ideal fluid model}


\author{Anna Karczewska} \email{A.Karczewska@wmie.uz.zgora.pl} \thanks{Corresponding author}
\affiliation{Faculty of Mathematics, Computer Science and Econometrics, University of Zielona G\'ora, Szafrana 4a, 65-516 Zielona G\'ora, Poland} 

\author{Piotr Rozmej} \email{P.Rozmej@if.uz.zgora.pl} 
\affiliation{Institute of Physics, 
Faculty of Physics and Astronomy,
University of Zielona G\'ora, Szafrana 4a, 65-516 Zielona G\'ora, Poland} 

\date{\today}

\begin{abstract}
We study the problem of gravity surface waves for the ideal fluid model in (2+1)-dimensional case. We apply a systematic procedure for deriving the Boussinesq equations for a prescribed relationship between the orders of four expansion parameters, the amplitude parameter $\alpha$, the long-wavelength parameter $\beta$, the transverse wavelength parameter $\gamma$, and the bottom variation parameter $\delta$. We also take into account surface tension effects when relevant. For all considered cases, the (2+1)-dimensional Boussinesq equations can not be reduced to a single nonlinear wave equation for surface elevation function. On the other hand, they can be reduced to a single, highly nonlinear partial differential equation for an auxiliary function $f(x,y,t)$ which determines the velocity potential but is not directly observed quantity. The solution $f$ of this equation, if known, determines the surface elevation function.  
We also show that limiting the obtained the Boussinesq equations to (1+1)-dimensions one recovers well-known cases of the KdV, extended KdV, fifth-order KdV, and Gardner equations. \\
\end{abstract}


\maketitle

\section{Introduction}\label{intro}

The Korteweg-de Vries equation \cite{KdV} is one of the most widely used nonlinear wave equations. It was derived by the perturbation calculus method from a model of an ideal fluid (nonviscous and incompressible) in which the motion is irrotational in (1+1)-dimensions and the bottom of the fluid container is flat. The simplest nonlinear wave equation in (2+1)-dimensions is the Kadomtsev-Petviashvili (KP for short) equation \cite{KP}. Both the KdV and KP equations are integral and satisfy many conservation laws. Their analytical solutions are also known.

Since nonlinear waves on the surface of seas and oceans are very important in practice, there are many studies in the literature on (2+1)-dimensional equations of the KdV or KP type, see, e.g.,  \cite{Wazwaz08,Wazwaz09,Peng10,Wazwaz11,WZZQ14,Adem16,Zhang17,Batwa18,Wanga19,Lou20}. The equations used in these studies are also integral, allowing their authors to construct analytical solutions of various types (solitons, periodic solutions, lumps, breathers).

The extension of the ideal fluid model to the (2+1)-dimensional case leads to significantly more complicated equations. The only such attempt known to us is the work   \cite{Fok}. Unfortunately, this particular paper is erroneous, as we demonstrated in \cite{RK21}. To our best knowledge, there are no correct (2+1)-dimensional studies that take into account the model of an ideal fluid in full detail. The reason is that the resulting equations are non-integrable.

In this paper, we generalise the perturbative approach described in \cite{BS2013} and \cite{KRcnsns} for (1+1)-dimensional case. The authors of \cite{BS2013} showed that for the flat bottom, the KdV, extended KdV, fifth-order KdV, and Gardner equations can be derived for different relations between small parameters determining nonlinearity $(\alpha)$ and dispersion $(\beta)$. In \cite{KRcnsns}, we generalised this approach for the case with an uneven bottom, assuming that the bottom variations are much smaller than the fluid's depth. Here, we follow \cite{KRcnsns} attempting to derive analogous wave equations for (2+1)-dimensional case. It turns out that while it is possible to derive a system of appropriate (2+1)-dimensional Boussinesq equations, it is rather intractable to reduce them to a wave equation on the surface of a fluid. However, it is possible to obtain a single nonlinear partial differential equation of the wave type in which the argument is an auxiliary function that determines the velocity potential. The solution of such an equation, if known, determines the (2+1)-dimensional function representing the time-dependent surface profile.

The paper is organised as follows. In section \ref{descr} we recall the model, define small parameters $\alpha,\beta,\gamma,\delta$ and set up the set of Euler's equations in scaled dimensionless variables. In section \ref{a1b1g1d1}, the case when all small parameters are of the same order is considered. For $\gamma=0$ and (1+1)-dimensions, this case leads to the Korteweg-de Vries equation extended 
for an uneven bottom, derived by us in \cite{KRcnsns}, and when the bottom is flat ($\delta=0$) reduces to the usual Korteweg-de Vries equation.
Section \ref{a1b1g1d2} is devoted to the case when $\alpha,\gamma$ are of the order of $\beta$, but $\delta$ is of the order of $\beta^{2}$. This case, for $\gamma=0$ and (1+1)-dimensions leads to the extended KdV equation generalised for an uneven bottom, derived in \cite{KRcnsns}, and when the bottom is flat ($\delta=0$) leads to the extended KdV equation derived by Marchant and Smyth in \cite{MS90}.
In section \ref{a2b1g2d2} we consider the case when parameters $\alpha,\gamma,\delta$ are of the order of $\beta^{2}$ and perturbation approach is extended to second order. Here, limitation to (1+1)-dimensions leads to fifth-order KdV equation generalised in \cite{KRcnsns} for an uneven bottom.
Section \ref{a1b2g2d2} deals with the case when $\alpha$ is the leading parameter, $\beta,\gamma,\delta$ are of the order or $\alpha^{2}$, and derivations are performed up to second order. In this case, limitation to (1+1)-dimensions leads to the Gardner equation extended 
for an uneven bottom \cite{KRcnsns}, and when $\delta=0$, that is, when the bottom is flat, reduces to the usual Gardner equation. Section \ref{concl} contains conclusions.


\section{Description of the model} \label{descr}
Let us consider the inviscid and incompressible fluid model whose motion is irrotational in a container with an impenetrable bottom. 
In dimensional variables, the set of hydrodynamical equations has the following form
\begin{eqnarray}  \label{g1}
\phi_{xx} + \phi_{yy} + \phi_{zz}&=& 0, \quad \mbox{in~the~whole~volume}, \\ \label{g2}
\phi_z - (\eta_x \phi_x + \eta_y \phi_y +\eta_t) &=& 0, \quad \mbox{at~the~surface},  \\ \label{g3}
\phi_t + \frac{1}{2}(\phi_x^2+\phi_y^2+\phi_z^2) +g\eta +\frac{p_{s}}{\varrho} &=& 0, \quad \mbox{at~the~surface}, \\ \label{g4}
\phi_z -h_{x}\phi_{x} -h_{y}\phi_{y} &=& 0, \quad \mbox{at~the~impenetrable~bottom}.
\end{eqnarray}
Here $\phi(x,y,z,t)$ denotes the velocity potential, $\eta(x,y,t)$ denotes the surface profile function, $g$ is the gravitational acceleration, $\varrho$ is fluid's density, 
and $p_{s}$ is additional pressure due to the surface tension~\cite{AN2010}
\begin{equation} \label{psurf}
 p_{s} =-T\,\nabla \cdot \left(\frac{\nabla \eta}{(1+\vert \nabla \eta\vert^2)^{1/2}} \right)= T \,\frac{  \left(1+\eta_{y}^{2}\right)\eta_{xx}+\left(1+\eta_{x}^{2}\right)\eta_{yy}-2\eta_{x}\eta_{y}\eta_{xy} }{ \left(1+\eta_{x}^{2}+\eta_{y}^{2}\right)^{3/2}} ,
\end{equation}
where $T$ is fluid's surface tension coefficient, and $\nabla=(\partial_x,\partial_y)$ is two-dimensional gradient.
The bottom can be non-flat and is described by the function $h(x,y)$. 
Indexes denote partial derivatives, i.e.\ $\phi_{xx}\equiv \frac{\partial^2 \phi}{\partial x^2}$, and so on. 

The next step consists in introducing a standard scaling to dimensionless variables (in general, it could be different in $x$-, $y$- and $z$-direction)
\begin{align} \label{bezw}
   \tilde{x}= & x/L, \quad \tilde{y}= y/l,\quad \tilde{z}= z/H, \quad \tilde{t}= t/(L/\sqrt{gH}), 
\quad \tilde{\eta}= \eta/A,\quad \tilde{\phi}= \phi /(L\frac{A}{H}\sqrt{gH}), 
\end{align}
where $A$ is the amplitude of surface distortions from equilibrium shape (flat surface), $H$ is average fluid depth, $L$ is the average wavelength (in $x$-direction), and $l$ is a wavelength in $y$-direction. In general,  $l$ should be the same order as $L$, but not necessarily equal. 
Then the set (\ref{g1})-(\ref{g4}) takes in scaled variables the following form (here and next, we omit the tilde signs)
\begin{eqnarray}  \label{G1}
\beta \phi_{xx} + \gamma\phi_{yy} + \phi_{zz}&=& 0,  \\ \label{G2}
 \eta_t +\alpha (\eta_x\phi_x+\frac{\gamma}{\beta}\eta_y\phi_y)-\frac{1}{\beta}\phi_z &=& 0,\quad \mbox{for} \quad z=1+\alpha\eta,\\ \label{G3}
\phi_t + \frac{1}{2}\alpha \left(\phi_x^2+\frac{\gamma}{\beta}\phi_y^2+\frac{1}{\beta}\phi_z^2\right) + \eta + \text{ST}
&=& 0,\quad \mbox{for} \quad z=1+\alpha\eta,  \\ \label{G4}
\phi_z - \beta\delta h_x\phi_x -\gamma\delta h_y\phi_y  &=& 0, \quad \mbox{for} \quad z=\delta h.
\end{eqnarray}
Besides standard small parameters $\alpha=\frac{a}{H}$,  $\beta=\left(\frac{H}{L}\right)^2$ and $\gamma=\left(\frac{H}{l}\right)^2$, which are sufficient for the flat bottom  case, we introduced another one defined as $\delta=\frac{a_h}{H}$, where $a_h$ denotes the amplitude of bottom variations \cite{KRcnsns,KRI}.
In the perturbation approach, all these parameters, $\alpha,\beta,\gamma,\delta$, are assumed to be small but not necessarily of the same order. ST is the term originating from surface tension. The explicit form of this term and its useful approximation up to second order is given below, in (\ref{stt}). For details, see the  Appendix \ref{appA}.
\begin{align} \label{stt}
\text{ST} & = -\tau  \,\frac{ \beta\left(1+\alpha^{2}\beta\eta_{y}^{2}\right)\eta_{xx}+\gamma\left(1+\alpha^{2}\beta\eta_{x}^{2}\right)\eta_{yy}-2\alpha^{2}\beta\gamma\eta_{x}\eta_{y}\eta_{xy} }{ \left(1+\alpha^{2}\beta\eta_{x}^{2}+\alpha^{2}\gamma\eta_{y}^{2}\right)^{3/2}}= -\tau\,(\beta\eta_{xx}+\gamma \eta_{yy} +O(\beta^4)).
\end{align}
The Bond number is defined as $\tau=\frac{T}{\varrho g H^{2}}$. For shallow-water waves, when $H\gtrsim$ 1 m, ST term can be safely neglected, since $\tau < 10^{-7}$, but it becomes important for waves in thin fluid layers. 
In sections \ref{a1b1g1d1} and \ref{a1b1g1d2} we neglect surface tension terms, in sections \ref{a2b1g2d2} and \ref{a1b2g2d2}
we take them into account.
 
The standard perturbation approach to the system of Euler's equations (\ref{G1})-(\ref{G4}) consists of the following steps. First, the velocity potential is sought in the form of power series in the vertical coordinate
\begin{equation} \label{Szer}
\phi(x,y,z,t)=\sum_{m=0}^\infty z^m\, \phi^{(m)} (x,y,t),
\end{equation}
where ~$\phi^{(m)} (x,y,t)$ are yet unknown functions. The Laplace equation (\ref{G1}) determines $\phi$ in the form which involves only two unknown functions with the lowest $m$-indexes, $f(x,y,t):=\phi^{(0)} (x,y,t)$ and $F(x,y,t):=\phi^{(1)} (x,y,t)$ and their space derivatives. Hence,
\begin{align} \label{Szer1}
\phi(x,y,z,t) & =\sum_{m=0}^\infty \frac{(-1)^m}{(2m)!} z^{2m}\, (\beta\partial_{xx}+\gamma\partial_{yy})^m f(x,y,t) \\ & 
 + \sum_{m=0}^\infty \frac{(-1)^m}{(2m+1)!} z^{2m+1}\,  (\beta\partial_{xx}+\gamma\partial_{yy})^{m} F(x,y,t). \nonumber
\end{align}
The explicit form of this velocity potential reads as
\begin{align} \label{pot8}
\phi = & \hspace{3ex} f -\frac{1}{2} z^2 (\beta f_{2x}+\gamma f_{2y}) + \frac{1}{24} z^4 (\beta^2 f_{4x}+2\beta\gamma f_{2x2y}+\gamma^2f_{4y}) 
+ \cdots  \nonumber \\ & 
+  z F-\frac{1}{6} z^3 (\beta F_{2x}+\gamma F_{2y})+ \frac{1}{120} z^5 (\beta^{2} F_{4x}+2\beta\gamma F_{2x2y}+\gamma^{2} F_{4y})+ \cdots 
\end{align}

Next, the boundary condition at the bottom (\ref{G4}) is utilized. For the flat bottom case, it implies $F=0$, simplifying substantially next steps. In particular, $F=0$  makes it possible to derive the Boussinesq equations up to arbitrary order. 
For an uneven bottom, the equation (\ref{G4}) determines a differential equation relating $F$ to $f$. This differential equation can be resolved to obtain $F(f,f_{x},f_{xx},h,h_{x})$ but this solution can be obtained only up to some particular order in leading small parameter. Then, the velocity potential is substituted into kinematic and dynamic boundary conditions at the unknown surface (\ref{G2})-(\ref{G3}). Retaining only terms up to a given order, one obtains the Boussinesq system of two equations for unknown functions $\eta,f$ valid only up to a given order in small parameters. The resulting equations, however, depend substantially on the ordering of small parameters.

In 2013, Burde and Sergyeyev \cite{BS2013} demonstrated that for the case of (1+1)-dimensional and the flat bottom, the KdV, the extended KdV, fifth-order KdV, and Gardner equations can be derived from the same set of Euler's equations (\ref{G1})-(\ref{G4}). Different final equations result from the different ordering of small parameters and consistent perturbation approach up to first or second order in small parameters.

 In 2020, we extended their results to cases with an uneven bottom in \cite{KRcnsns}, but still in (1+1)-dimensional theory.  We showed that the terms originating from the bottom have the same universal form for all these four nonlinear equations. However, the validity of the obtained generalised wave equations is limited to the cases when the bottom functions are piecewise linear. On the other hand, the corresponding sets of the Boussinesq equations are valid for the arbitrary form of the bottom functions. 

In the present paper there are four small parameters, $\alpha,\beta,\gamma,\delta$. In order to make calculations easier, we will follow the idea from \cite{BS2013,KRcnsns}, relating all small parameters to a single one, called \emph{leading  parametr}. This method allows us for easier control of the order of different terms, but the final forms of the obtained equations are presented in original parameters $\alpha,\beta,\gamma,\delta$. 
We discuss several cases, which are listed in Table \ref{tab1}. The table does  not contain all possible second-order cases, but only those that lead to well-known KdV-type and Gardner equations when reduced to (1+1)-dimensions.

\begin{table}[hbt] \caption{Different ordering of small parameters considered in the paper.} \label{tab1} \begin{center} 
\begin{tabular}{||c|c|c|c|c|c||} \hline Case & \hspace{4ex}$\alpha$\hspace{4ex} & \hspace{4ex}$\beta\hspace{4ex}$ & \hspace{2ex}$\gamma$ \hspace{2ex} &\hspace{2ex}$\delta$\hspace{2ex} & (1+1)-dim, $\delta=0$ \\  \hline 
1 & $O(\beta)$ & leading   & $O(\beta)$ & $O(\beta)$  &  KdV\\ \hline 
2 & $O(\beta)$ &  leading  & $O(\beta)$& $O(\beta^2)$  & extended KdV\\ \hline 
3 & $O(\beta^2)$ & leading   &$O(\beta^{2})$ & $O(\beta^2)$  &  5th-ord KdV\\ \hline 
3 & leading  & $O(\alpha^2)$  & $O(\alpha^{2})$ & $O(\alpha^{2})$  & Gardner \\ \hline 
\end{tabular} \end{center} 
\end{table}  

\section{Case~1: \hspace{1ex}  $\bal=O(\bbe)$, \hspace{1ex} $\bga=O(\bbe)$, \hspace{1ex} $\bde=O(\bbe) $} \label{a1b1g1d1}

Let us begin with the assumption that all small parameters are of the same order. For easier control of the orders of different terms let us denote
$$ \alpha= A\,\beta, \quad \gamma = G\,\beta, \quad \delta= D\,\beta, $$
where $A,G,D$ are arbitrary but close to 1.

Insertion the velocity potential (\ref{pot8}) into the boundary condition at the bottom (\ref{G4}) imposes the following relation
\begin{align} \label{r9}
F & = \beta^{2} D \left(h_x f_x +G h_y f_y + h(f_{xx}+G f_{yy})\right) \\ & \hspace{2ex} + \beta^{3}D^{2}\left(h h_x F_x +G h h_y F_y+\frac{1}{2} h^2 (F_{xx}+G F_{yy})\right) +O(\beta^{5}).   \nonumber \end{align}
Equation (\ref{r9}), when only terms up to second order in small parameters are retained, that is 
\begin{equation} \label{r9a}
F =\beta^{2} D \left(h_x f_x + G h_y f_y +h(f_{xx}+ G f_{yy})\right)
\end{equation}
allows us to express $F$ by $f$ and reduce the set of unknown functions to only $\eta$ and $f$.
In higher order approxiation equation (\ref{r9}) cannot be resolved for $F$.
Then, retaining only terms up to second order, one obtains the velocity potential, valid up to second-order approximation, expressed in terms of a single unknown function $f(x,y,t)$ as
\begin{align} \label{vp2}
\phi  = f & -\frac{1}{2}\beta\, z^2 (f_{xx}+G f_{yy}) + \frac{1}{24}\beta^2\, z^4 (f_{xxxx}+2 G f_{xxyy}+G^2f_{yyyy})  \\ &  \nonumber
+ \beta^2 z\,D \left(h_x f_x + G h_y f_y +h(f_{xx}+ G f_{yy})\right).
\end{align}

Now, we substitute the velocity potential (\ref{vp2}) into (\ref{G2}) and  (\ref{G3}), keeping terms only up to first order. Due to the term $\frac{1}{\beta}\phi_z$ in (\ref{G2}) and only second order valid  expression for the potential (\ref{vp2}) we can obtain valid Boussinesq's equations only in first order in  $\alpha,\beta,\gamma,\delta $. 

From (\ref{G2}) and (\ref{G3}) we get respectively 
\begin{align} \label{BR7}
\eta_t + f_{xx}+\frac{\gamma}{\beta}f_{yy} &
 +\alpha \left(\left(\eta f_x\right)_x+ \frac{\gamma}{\beta}\left(\eta f_y\right)_y \right)   
-\frac{1}{6}\,\beta \left(f_{4x}+2 \frac{\gamma}{\beta} f_{2x2y} +\left(\frac{\gamma}{\beta}\right)^2f_{4y}\right) \nonumber \\ & 
-\delta \left(\left( h f_x\right)_x +\frac{\gamma}{\beta} \left(h f_y\right)_y\right) = 0  \qquad  \mbox{and}\\  \label{BR8}
\eta +f_{t} & +\frac{1}{2} \alpha \left(f_{x}^{2} +\frac{\gamma}{\beta} f_{y}^{2}  \right)  -\frac{1}{2}\,\beta \left(f_{xxt} +\frac{\gamma}{\beta} f_{yyt}\right) =0.
\end{align}
Equations (\ref{BR7})-(\ref{BR8}) are the first order Boussinesq equations for (2+1)-dimensional shallow water problem with an uneven bottom. 

Let us check whether the reduction of the equations (\ref{BR7})-(\ref{BR8}) to (1+1)-dimensions leads to correct results. 
When we reduce these equations to (1+1)-dimensional (assuming translational invariance in $y$ direction, that is, setting all $y$-derivatives equal zero), then we arrive at 
\begin{align} \label{BR7-1}
\eta_t + f_{xx} & +\alpha \left(\eta f_x\right)_x   
-\frac{1}{6}\,\beta f_{4x}  -\delta \left( h f_x\right)_x  = 0, \\  \label{BR8-1}
\eta +f_{t}&+\frac{1}{2}\alpha f_{x}^{2} -\frac{1}{2}\,\beta f_{xxt} =0.
\end{align}
Denoting $w=f_x$ and taking $x$-derivative of (\ref{BR8-1}) one gets this set of Boussinesq's equation in the following form
\begin{align} \label{BR7-2}
\eta_t + w_{x} & +\alpha \left(\eta w\right)_x   
-\frac{1}{6}\,\beta w_{xxx}  -\delta \left( h w\right)_x  = 0, \\  \label{BR8-2}
w_{t} + \eta_x &+\alpha w w_{x} -\frac{1}{2}\,\beta w_{xxt} =0,
\end{align}
identical to the equations (17)-(18) derived by us in \cite{KRcnsns}. 
In \cite{KRcnsns}, we also demonstrated that these two equations could be made compatible and reduced to the KdV equation generalised for uneven bottom \cite[Eq.~(29)]{KRcnsns}, when the bottom profile $h(x)$ is a piecewise linear function. This equation has the following form
\begin{equation}\label{kdvg}
\eta_{t}+\eta_{x}+\frac{3}{2}\alpha\eta\eta_{x}+\frac{1}{6}\eta_{xxx}-\frac{1}{4}\delta(2h\eta_{x}+h_{x}\eta) =0.
\end{equation}
For the flat bottom, $\delta=0$, the equation (\ref{kdvg}) reduces to the usual Korteweg-de Vries equation.

In \cite{Fok}, the authors presented the derivation of two new (2+1)-dimensional third- and fifth-order nonlinear evolution equations  
when $\alpha,\beta,\gamma$ are of the same order and the bottom is flat ($\delta=0$). However, as we proved in \cite{RK21}, all results shown in \cite{Fok} are false since the derivation is inconsistent and violates the fundamental property of the velocity potential. When the method used by the authors is applied consistently, the problem is reduced to well known KdV equation. For details, see \cite{RK21}.

The Boussinesq equations (\ref{BR7})-(\ref{BR8}) look very complicated. It is hard to imagine how one can get a single equation for wave profile  from them. On the other hand, one can reduce this set to a single equation by inserting $\eta =-\left[ f_{t} +\frac{1}{2} \alpha \left(f_{x}^{2} +\frac{\gamma}{\beta} f_{y}^{2}  \right)  -\frac{1}{2}\,\beta \left(f_{xxt} +\frac{\gamma}{\beta} f_{yyt}\right)\right]$ 
determined from (\ref{BR8}) into (\ref{BR7})
and retaining only terms up to first order. 
The result is
\begin{align} \label{rff}
f_{xx} + \frac{\gamma}{\beta} f_{yy} -f_{tt} & -\alpha\left[f_t \left(f_{xx} + \frac{\gamma}{\beta} f_{yy} \right)+ \left(f_x^2 +\frac{\gamma}{\beta} f_y^2\!\right)_t\right] \nonumber \\ & - \beta 
\left[\frac{1}{6} \left(f_{4x}+2\frac{\gamma}{\beta} f_{2x2y} +\left(\frac{\gamma}{\beta}\right)^2 f_{4y}\right) -\frac{1}{2}\left(f_{xxtt}+ \frac{\gamma}{\beta}f_{yytt} \right)\right]  \\ & -\delta \left[\left(h f_x\right)_x+ \frac{\gamma}{\beta}\left(h f_y\right)_y\right]=0. \nonumber
\end{align} 
Equation (\ref{rff}) can be simplified utilizing zeroth-order solutions.
In zeroth-order we have from (\ref{BR7})-(\ref{BR8})
\begin{equation} \label{0ord}
\eta_t + f_{xx}+\frac{\gamma}{\beta}f_{yy} =0 \qquad\quad \mbox{and} \qquad \quad\eta +f_{t}=0.
\end{equation}
So, $f_{t}=-\eta~$ and $f_{tt}=-\eta_{t}=f_{xx}+\frac{\gamma}{\beta} f_{yy}$. Then we have
$$f_{xxtt}= \frac{\partial^2}{\partial x^2} \left(f_{xx}+\frac{\gamma}{\beta} f_{yy}\right)= f_{4x}+ \frac{\gamma}{\beta} f_{xxyy},  \quad f_{yytt}= \frac{\partial^2}{\partial y^2} \left(f_{xx}+\frac{\gamma}{\beta} f_{yy}\right)= f_{xxyy}+ \frac{\gamma}{\beta} f_{4y},$$
and the first-order term 
$$ -\frac{1}{2}\left(f_{xxtt}+ \frac{\gamma}{\beta}f_{yytt} \right) =
 -\frac{1}{2}\left(f_{4x}+2\frac{\gamma}{\beta} f_{2x2y} +\left(\frac{\gamma}{\beta}\right)^2 f_{4y}\right)$$
receives the same form like the term with the factor $\frac{1}{6}$ in front.
Then equation (\ref{rff}) receives the simpler form
\begin{align} \label{rFF}
f_{xx} + \frac{\gamma}{\beta} f_{yy} -f_{tt} & -\alpha\left[f_t \left(f_{xx} + \frac{\gamma}{\beta} f_{yy} \right)+ \left(f_x^2 +\frac{\gamma}{\beta} f_y^2\!\right)_t\right] \nonumber \\ & + \beta 
\left[\frac{1}{3} \left(f_{4x}+2\frac{\gamma}{\beta} f_{2x2y} +\left(\frac{\gamma}{\beta}\right)^2 f_{4y}\right) \right]  -\delta \left[\left(h f_x\right)_x+ \frac{\gamma}{\beta}\left(h f_y\right)_y\right]=0. 
\end{align} 
If the solution $f(x,y,t)$ to (\ref{rFF}) is known, the equation (\ref{BR8}) supplies the surface profile function 
\begin{equation} \label{eFF}
\eta(x,y,t)=-\left[ f_{t} +\frac{1}{2} \alpha \left(f_{x}^{2} +\frac{\gamma}{\beta} f_{y}^{2}  \right)  -\frac{1}{2}\,\beta \left(f_{xxt} +\frac{\gamma}{\beta} f_{yyt}\right)\right].
\end{equation}

A preliminary idea how to look for solution to the equation (\ref{rFF}) is presented in the appendix \ref{appB}.



\section{Case~2: \hspace{1ex}  $\bal=O(\bbe)$, \hspace{1ex} $\bga=O(\bbe)$, \hspace{1ex} $\bde=O(\bbe^2)$} \label{a1b1g1d2}

Denote now
$$ \alpha= A\,\beta, \quad \gamma = G\,\beta, \quad \delta= D\,\beta^2.$$
In this case, the boundary condition at the bottom (\ref{G4}) imposes the following relation
\begin{align} \label{r9c3}
F & = \beta^{3} D\left[ (h f_x)_x+ G (h f_y)_y\right] + \frac{1}{2}\beta^{5} D^{2} \left[(h^2 F_x)_x+G(h^2 F_y)_y \right] +O(\beta^7),
\end{align}
which gives $F = \beta^{3} D \left[(h f_x)_x+ G (h f_y)_y\right]$ valid up to fourth order. Therefore, in this case, it is possible to obtain the Boussinesq equations valid up to third order in small parameters. 
  
From (\ref{G2}) we get (up to second order)
\begin{align} \label{r7c3}
\eta_{t} +f_{xx}+ \frac{\gamma}{\beta} f_{yy} &+\alpha\left[(\eta f_x)_{x}
+ \frac{\gamma}{\beta} (\eta f_y)_{y}\right] 
-\frac{1}{6} \beta\left[f_{4x}+2\frac{\gamma}{\beta}f_{2x2y} +\left(\frac{\gamma}{\beta}\right)^2 f_{4y}  \right] \nonumber \\ &  
- \frac{1}{2}\,\alpha\beta(\eta f_{3x})_{x}  - \frac{1}{2}\,\alpha\gamma \left[\eta_{x} f_{x2y}+2\eta f_{2x2y}+\eta_{y} f_{2xy} \right] - \frac{1}{120}\,\beta^2 f_{6x} \\ & -\frac{1}{40}\,\beta\gamma \left[f_{4x2y}+\frac{\gamma}{\beta}\,f_{2x4y}+\frac{1}{3} \left(\frac{\gamma}{\beta}\right)^{2}\! f_{6y} \right]  - \delta \left[ (h f_x)_x + \frac{\gamma}{\beta} (h f_y)_y\right]
=0,  \nonumber
\end{align}
whereas the result from  (\ref{G3}), up to second order, is
\begin{align} \label{r8c3}
\eta + f_t & + \frac{1}{2}\alpha \left(f_x^2 +\frac{\gamma}{\beta} f_y^2\right)-\frac{1}{2} \beta \left(f_{xxt} +  \frac{\gamma}{\beta}f_{yyt}\right) +\alpha\beta\left[\frac{1}{2} (f_{xx}^2 -f_x f_{3x})-\eta f_{xxt}\right] \nonumber \\ & +\alpha\gamma \left[\eta f_{yyt}+f_{xx} f_{yy} -\frac{1}{2} (f_x f_{xyy}+f_y f_{xxy})+ \frac{\gamma}{\beta}(f_{yy}^2-f_y f_{3y} )\right] \\ & +\frac{1}{24} \beta^2 \left[f_{4xt} + 2 \frac{\gamma}{\beta} f_{2x2yt} +  \left(\frac{\gamma}{\beta}\right)^{2}\! f_{4yt}\right] =0. \nonumber 
\end{align}
Note that the term  with $\delta$ in (\ref{r7c3}), originating from the uneven bottom, has identical form as such term in (\ref{BR7}) although now it is of second order.
Let us rewrite (\ref{r8c3}) in the form
\begin{align} \label{r8c3et}
\eta\left(1\!-\alpha\beta f_{xxt}-\!\alpha\gamma f_{yyt} \right) =-&\left\{ 
f_t + \frac{1}{2}\alpha\! \left(\! f_x^2 +\frac{\gamma}{\beta} f_y^2\right)-\frac{1}{2} \beta\! \left(\! f_{xxt} +  \frac{\gamma}{\beta}f_{yyt}\right) +\alpha\beta\left[\frac{1}{2} (f_{xx}^2 -f_x f_{3x})\right]\right.  \nonumber \\ & +\alpha\gamma \left[f_{xx} f_{yy} -\frac{1}{2} (f_x f_{xyy}+f_y f_{xxy})+ \frac{\gamma}{\beta}(f_{yy}^2-f_y f_{3y} )\right] \\ &  \left. +\frac{1}{24} \beta^2 \left[f_{4xt} + 2 \frac{\gamma}{\beta} f_{2x2yt} +  \left(\frac{\gamma}{\beta}\right)^{2}\! f_{4yt}\right] \right\}. \nonumber 
\end{align}
Multiplying both sides by $\left(1+\alpha\beta f_{xxt}+\alpha\gamma f_{yyt} \right)$ and retaining only terms up to second order we obtain
\begin{align} \label{r8c3eta} 
\eta =-& \left\{ f_t + \frac{1}{2}\alpha \left(f_x^2 +\frac{\gamma}{\beta} f_y^2\right)-\frac{1}{2} \beta \left(\!f_{xxt} +  \frac{\gamma}{\beta}f_{yyt}\!\right) +\alpha\beta\left[\frac{1}{2} (f_{xx}^2 -f_x f_{3x})\right]\right.  \nonumber \\ & +\alpha\gamma \left[f_{xx} f_{yy} -\frac{1}{2} (f_x f_{xyy}+f_y f_{xxy})+ \frac{\gamma}{\beta}(f_{yy}^2-f_y f_{3y} )\right] \\ &  \left. +\frac{1}{24} \beta^2 \left[f_{4xt} + 2 \frac{\gamma}{\beta} f_{2x2yt} +  \left(\frac{\gamma}{\beta}\right)^{2}\! f_{4yt}\right] \right\} + f_t\left( \alpha\beta f_{xxt}+\alpha\gamma f_{yyt} \right) . \nonumber 
\end{align}

Substituting $\eta$ defined by (\ref{r8c3eta}) into (\ref{r7c3}) and keeping only the terms up to second order allows us to obtain a nonlinear (2+1)-dimensional differential equation for the function $f(x,y,t)$. This equation has the form (\ref{r7-8fin}). 
In (\ref{r7-8fin}) we didn't come back to original parameters $\alpha,\beta,\gamma,\delta$ for easier recognition of first order and second order terms. 

\begin{align} 
&  f_{xx} +G f_{yy} -f_{tt} -A G (f_{tt} f_{yyt} + f_{t} f_{2y2t})
   -A (f_{tt} f_{2xt} + f_{t} f_{2x2t}) \nonumber \\ &
+ \beta \bigg[ 
   - G A^2 \left( f_{yyt} f_{x} f_{xt} + f_{t} f_{x} f_{x2yt} A^2
   + f_{t} f_{yyt} f_{xx} + f_{y} f_{xy} f_{2xt} 
   + f_{t} f_{yy} f_{2xt} - f_{t} f_{y} f_{2xyt} + f_{t} f_{yy}
  \right) \nonumber  
\\ & 
   \hspace{5ex} - G^2 A^2 \left( f_{y} f_{xy} f_{yyt} 
   + f_{t} f_{yy} f_{yyt}  + f_{t} f_{y} f_{3yt}  \right) 
 - A^2 \left( f_{x} f_{xt} f_{2xt} -f_{t} f_{xx} f_{2xt} - f_{t} f_{x} f_{3xt} \right)  \nonumber \\ &  
   \hspace{5ex} -2 G A f_{y} f_{xy}  -2 f_{x} f_{xt} - A f_{t} f_{xx}
   +\frac{1}{2} G f_{2y2t}
   -\frac{1}{6} G^2 f_{4y} 
   +\frac{1}{2} f_{2x2t}
   -\frac{1}{3} G f_{2x2y}
   -\frac{1}{6} f_{4x}  \bigg] 
\nonumber \\ & 
+ \beta^2 \left[ A^2 G^3 \left( f_{xy} f_{yyt} f_{3y} + f_{t} f_{3y} f_{3yt} 
+ f_{t} f_{yyt} f_{4y} \right) +\frac{1}{120} G^3 f_{6y}   \right. \nonumber \\ &  
\hspace{5ex} + A^2 G^2 \left(
   -\frac{3}{2}  f_{y}^2 f_{yy} 
   +\frac{1}{2}  f_{yyt} f_{xt} f_{x2y} 
   +\frac{1}{2}  f_{t} f_{x2y} f_{x2yt}  
   +\frac{1}{2}  f_{xy} f_{3y} f_{2xt} 
   +\frac{1}{2}  f_{t} f_{4y} f_{2xt}   \right.
\nonumber \\ & \left. 
 \hspace{5ex}  +\frac{1}{2}  f_{xy} f_{yyt} f_{2xy} 
   +\frac{1}{2}  f_{t} f_{3yt} f_{2xy} 
   +\frac{1}{2}  f_{t} f_{3y} f_{2xyt} 
   + f_{t} f_{yyt} f_{2x2y}   \right) \nonumber  \\ & 
 \hspace{5ex}  
+ A G^2 \left(  -\frac{1}{2}  f_{yy} f_{yyt}  + f_{xy} f_{3y}  
   + f_{y} f_{3yt} +\frac{1}{2}  f_{t} f_{4y}  \right) \label{r7-8fin}  \\ & 
\hspace{5ex} + A^2 G \left(
   -\frac{1}{2}  f_{yy} f_{x}^2 
   -\frac{1}{2}  f_{y}^2 f_{xx} 
   +\frac{1}{2}  f_{xt} f_{x2y} f_{2xt} 
   +\frac{1}{2}  f_{xy} f_{2xt} f_{2xy} 
   +\frac{1}{2}  f_{t} f_{2xy} f_{2xyt} 
   +\frac{1}{2}  f_{yyt} f_{xt} f_{3x} \right.
\nonumber \\ & \left. 
  \hspace{12ex}  +\frac{1}{2}  f_{t} f_{x2yt} f_{3x} 
   +\frac{1}{2}  f_{t} f_{x2y} f_{3xt} 
   +\frac{1}{2}  f_{t} f_{yyt} f_{4x} 
   + f_{t} f_{2xt} f_{2x2y}
   -2 f_{y} f_{x} f_{xy}         \right) \nonumber \\ & 
 \hspace{5ex} + G^2  \left(
   -\frac{1}{24} f_{4y2t}
   +\frac{1}{40} f_{2x4y}      \right)
+ D G \left(- h_{h} f_{y} - h f_{yy} \right) - D \left( h f_{x} + h f_{xx}    \right) \nonumber \\ & 
 \hspace{5ex}+ A G \left(
   + f_{xt} f_{x2y}  
   + f_{x} f_{x2yt} 
   -\frac{1}{2}  f_{yyt} f_{xx} 
   -\frac{1}{2}  f_{yy} f_{2xt} 
   + f_{xy} f_{2xy} 
   + f_{y} f_{2xyt} 
   + f_{t} f_{2x2y}      \right)  \nonumber \\ & 
 \hspace{5ex}+ G \left(
   -\frac{1}{12} f_{2x2y2t} 
   +\frac{1}{40} f_{4x2y}      \right)  
+ A \left(
   -\frac{1}{2} f_{xx} f_{2xt}
   + f_{xt} f_{3x}
   + f_{x} f_{3xt}
   +\frac{1}{2} f_{t} f_{4x}     \right)  \nonumber \\ & \left.
 \hspace{5ex}+ A^2 \left(
    \frac{1}{2}  f_{xt} f_{2xt} f_{3x}
   +\frac{1}{2}  f_{t}  f_{3x} f_{3xt}
   +\frac{1}{2}  f_{t} f_{2xt} f_{4x}     \right) 
   -\frac{1}{24} f_{4x2t}
   +\frac{1}{120} f_{6x}
 \right] \nonumber=0. \nonumber
\end{align}
Equation (\ref{r7-8fin}) is the single nonlinear wave equation for auxiliary function $f$ determining the velocity potential, second order in small parameters for the case $~\alpha=O(\beta), ~\gamma=O(\beta), ~\delta=O(\beta^2)$.

In principle, if the solution $f(x,y,t)$ to (\ref{r7-8fin}) is known, the equation (\ref{r8c3eta}) supplies the surface profile function. The complexity of equations (\ref{r8c3eta}) and (\ref{r7-8fin}) may be the reason why there are so many small amplitude wrinkles and ripples that are observed on the surface of seas and oceans.
However, the complexity of the equation (\ref{r7-8fin}) gives a little hope for finding the solution.

Let us check how the Boussinesq equations (\ref{r7c3})-(\ref{r8c3}) look like when they are reduced to (1+1)-dimensions.
From  (\ref{r7c3}) we get (denoting $w=f_{x}$)
\begin{align} \label{r7c3-1d}
\eta_{t} +w_{x} &+\alpha (\eta w)_{x}-\frac{1}{6} \beta w_{3x}  
- \frac{1}{2}\,\alpha\beta(\eta w_{2x})_{x}  - \frac{1}{120}\,\beta^2 w_{5x}  
 - \delta (h w)_x =0,  
\end{align}
whereas from (\ref{r8c3}) after differentiation over $x$ we obtain
\begin{align} \label{r8c3-1d}
\eta_{x} + w_t & + \alpha w w_{x} - \frac{1}{2} \beta w_{2xt}-\alpha\beta\left[\frac{1}{2}\left(w_{x}w_{2x}-w w_{3x}\right) -(\eta w_{xt})_{x}\right] +\frac{1}{24} \beta^2 w_{4xt}  =0.
\end{align}
The (1+1)-dimensional Boussinesq equations (\ref{r7c3-1d})-(\ref{r8c3-1d}) are identical with  \cite[Eqs.~(37)-(38)]{KRcnsns}, and for the case of flat bottom ($\delta=0$) are identical with  \cite[Eqs.~(11)-(12)]{BS2013}. In the latter case these equations lead to the extended KdV equation \cite{MS90}. 
So, the full (2+1)- dimensional Boussinesq equations (\ref{r7c3})-(\ref{r8c3}) seem to be correct.

\section{Case~3: \hspace{1ex}  $\bal=O(\bbe^2)$, \hspace{1ex} $\bga=O(\bbe^2)$, \hspace{1ex} $\bde=O(\bbe^2)$} \label{a2b1g2d2}

Denote now
$$ \alpha= A\,\beta^2, \quad \gamma = G\,\beta^2, \quad \delta= D\,\beta^2.$$
Now, the boundary condition at the bottom (\ref{G4}) imposes the following relation
\begin{align} \label{r9c5}
F & = \beta^{3} D (h f_x)_x +\beta^4 D G (h f_y)_y -\beta^{5}\frac{1}{2} D (h^2 F_x)_x + O(\beta^6),
\end{align}
which gives $F = \beta^{3} D \left[(h f_x)_x\right]$ and consequently the velocity potential valid up to fourth order. Therefore, in this case it is possible to obtain the Boussinesq equations valid up to third order in small parameters. We will proceed up to second order which is enough complicated.
From (\ref{G2}) we obtain
\begin{align} \label{r7c5}
\eta_t +f_{xx} & -\frac{1}{6}\beta f_{4x} + \frac{\gamma}{\beta} f_{yy} +\alpha (\eta f_x)_x-\frac{1}{120}\beta^2 f_{6x} - \frac{1}{3}\gamma f_{2x2y}
-\delta (h f_x)_x =0, 
\end{align}
whereas the result from (\ref{G3}) is 
\begin{align} \label{r8c5}
\eta+ f_t -\frac{1}{2}\beta f_{xxt} +\frac{1}{2}\alpha f_x^2 +\frac{1}{24} \beta^2 f_{4xt} - \frac{1}{2}\gamma f_{yyt} =0.
\end{align}
In (\ref{r8c5}), we didn't take into account terms from surface tension (setting $\tau = 0$ in (\ref{G3})). This omission is fully justified when water depth is on the order of meters.
Now, $\eta$ determined from (\ref{r8c5}) is substituted into  (\ref{r7c5}). Limiting this equation up to second order one obtains
\begin{align} \label{r78c5}
f_{xx} -f_{tt} & +\frac{\gamma}{\beta} f_{yy}  + \beta \left( \frac{1}{2}f_{xxtt}  -\frac{1}{6} f_{4x}  \right) -\beta^2\left(\frac{1}{24} f_{4x2t} + \frac{1}{120} f_{6x}\right) \\ & +\gamma\left(
-\frac{1}{3}f_{xxyy} +\frac{1}{2}f_{yytt} \right)
 -\alpha(2f_x f_{xt} +f_t f_{xx}) -\delta(hf_x)_x = 0. \nonumber 
\end{align}
Equation (\ref{r78c5}) is the single nonlinear wave equation for auxiliary function $f$ determining the velocity potential, second order in small parameters for the case $~\alpha=O(\beta^2), ~\gamma=O(\beta^2), ~\delta=O(\beta^2)$.

For a check, let us reduce the Boussinesq equations (\ref{r7c5})-(\ref{r8c5}) to (1+1)-dimensional case setting $\gamma =0$. Then, after differentiating (\ref{r8c5}) over $x$ and denoting $w=f_x$ we obtain the set
\begin{align} \label{r7c5d1}
\eta_t +w_{x} & -\frac{1}{6}\beta w_{3x} +\alpha (\eta w)_x-\frac{1}{120}\beta^2 w_{5x} -\delta (h f_x)_x =0  \qquad \mbox{and}  \\  \label{r8c5d1}
w_t +\eta_x & -\frac{1}{2}\beta w_{xxt} +\alpha w w_x+\frac{1}{24} \beta^2 w_{4xt} =0.
\end{align}
This set of equations, for  $\delta=0$, is identical to the equations (29)-30) in \cite{BS2013}, limited to second order. Eliminating by the standard method the function $w$, one obtains from them the well-known fifth-order KdV equation when $\delta=0$, i.e., for the flat bottom. When $\delta\ne 0$, the final wave equation is the fifth-order KdV generalised for the uneven bottom \cite[Eq.~(68)]{KRcnsns}, in which $\tau =0$.

\subsection{Case~5a: \hspace{1ex}  $\bal=O(\bbe^2)$, \hspace{1ex} $\bga=O(\bbe^2)$, \hspace{1ex} $\bde=O(\bbe^2)$, \hspace{1ex} $\boldsymbol{\tau}\ne 0$ \hspace{20ex}}   

For thin fluid layers, surface tension may introduce significant changes in the dynamics. Using surface tension term in the form (\ref{stt}) in dynamical boundary condition at the surface (\ref{G3}), that is
\begin{align}\label{G3st}
\phi_t + \frac{1}{2}\alpha \left(\phi_x^2+\frac{\gamma}{\beta}\phi_y^2+\frac{1}{\beta}\phi_z^2\right) + \eta -\tau\, (\beta\eta_{xx}+\gamma \eta_{yy}  )
&= 0,\quad \mbox{for} \quad z=1+\alpha\eta,
\end{align}
we obtain instead (\ref{r8c5}) more complicated equation
\begin{align} \label{r8c5a}
\eta+ f_t -\beta\left(\frac{1}{2} f_{xxt} +\tau \eta_{xx}\right)+\frac{1}{2}\alpha f_x^2 +\frac{1}{24} \beta^2 f_{4xt} -\gamma \left(\frac{1}{2}f_{yyt} +\tau \eta_{yy}\right) =0.  
\end{align}
In this case, because we can not express $\eta$ trough only $f$ and its derivatives, 
we can not reduce the Boussinesq equations (\ref{r7c5})-(\ref{r8c5a}) to a single equation for the function $f$. However, limiting to (1+1)-dimensions (as usual setting $\gamma =0$, $w=f_{x}$ and differentiating (\ref{r8c5a}) over $x$) we obtain
\begin{align} \label{r7c5d1a}
\eta_t +w_{x} & -\frac{1}{6}\beta w_{3x} +\alpha (\eta w)_x-\frac{1}{120}\beta^2 w_{5x} -\delta (h f_x)_x =0  \qquad \mbox{and}  \\  \label{r8c5d1a}
w_t +\eta_x & -\beta\left(\frac{1}{2} w_{xxt}+\tau \eta_{xxx}\right)  +\alpha w w_x+\frac{1}{24} \beta^2 w_{4xt} =0.
\end{align}
These equations are identical to \cite[Eqs.~(61)-(62)]{KRcnsns}, which lead to the  fifth-order KdV equation generalised for an uneven bottom \cite[Eq.~(68)]{KRcnsns}. 

\section{Case~4: \hspace{1ex}  $\bbe=O(\bal^2)$, \hspace{1ex} $\bga=O(\bal^2)$, \hspace{1ex} $\bde=O(\bal^2)$} \label{a1b2g2d2}

Denote now
$$ \beta= B\,\alpha^2, \quad \gamma = G\,\alpha^2, \quad \delta= D\,\alpha^2.$$
Now, the boundary condition at the bottom (\ref{G4}) imposes the following relation
\begin{align} \label{r9c6}
F & = \alpha^4 (B D(h f_x)_x+ D G (h f_y)_y) +\alpha^{6}\frac{1}{2}(B D^2 (h^2 F_x)_x + G D^2  (h f_y)_y) + O(\alpha^8),
\end{align}
which gives $F =\alpha^4 (B D(h f_y)_y+ D G (h f_y)_y) $ and consequently the velocity potential valid up to fourth order in the leading parameter. Therefore, in this case, it is possible to obtain the Boussinesq equations valid up to third order in small parameters. We will proceed up to the second order, which is complicated enough. Then we can safely neglect higher order terms in surface tension term (\ref{stt}) and use the dynamic boundary condition  
in the  form (\ref{G3st}).
From kinematic boundary condition at the surface (\ref{G2}), we obtain
\begin{align} \label{B1c6}
\eta_t +f_{xx} +\frac{\gamma}{\beta} f_{yy} & +\alpha
\left((\eta f_x)_x +\frac{\gamma}{\beta}(\eta f_y)_y \right)-\frac{1}{6}\beta \left( f_{4x}+\frac{\gamma^{2}}{\beta^{2}} f_{4y}\right) \nonumber \\ & -\frac{1}{3}\gamma f_{2x2y} -\delta\left( (h f_x)_x +\frac{\gamma}{\beta}(h f_y)_y \right)=0,
\end{align}
whereas, from the dynamic boundary condition at the surface (\ref{G3st}), the result is
\begin{align} \label{B2c6}
\eta +f_t +\frac{1}{2}\alpha\left(f_x^2+ \frac{\gamma}{\beta} f_y^2\right) 
-\frac{1}{2} (\beta f_{xxt}+\gamma f_{yyt})-\tau (\beta \eta_{xx}+\gamma \eta_{yy}) =0.
\end{align} 
Equations (\ref{B1c6})-(\ref{B2c6}) are the Boussinesq	equations for the case when $\beta\approx\gamma\approx\delta =O(\alpha^{2})$.

When $\tau$ is of the order of 1, what occurs for thin fluid layers, equation (\ref{B2c6}) 
can not be resolved to obtain $\eta=G(f,f_{x},f_{y},f_{{t}},\ldots)$, that is, as a function  of $f$ and its derivatives only. Therefore, in this case, the Boussinesq set (\ref{B1c6})-(\ref{B2c6}) can not be reduced to a single equation for $f$ function.

Limiting equations (\ref{B1c6})-(\ref{B2c6}) to (1+1)-dimnesions by setting
all $y$-derivatives equal to zero (alternatively setting $\gamma=0$) we obtain from (\ref{B1c6})
\begin{equation} \label{B1c61}
\eta_t +w_{x}  +\alpha (\eta w)_x -\frac{1}{6} \beta w_{xxx} -\delta
(h w)_x =0.
\end{equation}
From (\ref{B2c6}), after setting $\gamma=0$ and differentiating over $x$ we have
\begin{equation} \label{B2c61}
\eta_x +w_{t}  + \alpha w w_x -\frac{1}{2} \beta w_{xxt} -\tau \beta \eta_{xxx}=0.
\end{equation}
Equations (\ref{B1c61})-(\ref{B2c61}) are identical with equations (85)-(86) from \cite{KRcnsns}, when the latter are limited to second order. Therefore, one can make them compatible and derive a single wave equation for wave profile function $\eta(x,t)$. Such procedure leads to the equation identical with the equation (91) from \cite{KRcnsns}, which is the Gardner equation generalised for an uneven bottom. For $\delta=0$, that is for the flat bottom,  equation (\ref{B1c61}) receives the form \cite[Eq.~(92)]{KRcnsns}, the usual Gardner equation.

The reduction of the equations (\ref{B1c6})-(\ref{B2c6}) to the equation for a single unknown function $f$  is possible for cases when the depth of the fluid is of the order of meters. Then $\tau$ is so small that terms $-\tau (\beta \eta_{xx}+\gamma \eta_{yy})$ in (\ref{B2c6}) can be safely neglected. In such case substituting $\eta = -\left[f_t +\frac{1}{2} \left(f_x^2+ \frac{\gamma}{\beta} f_y^2\right) -\frac{1}{2} (\beta f_{xxt}+\gamma f_{yyt})\right]$ obtained from (\ref{B2c6}) with $\tau=0$ into (\ref{B1c6}) we obtain a single nonlinear wave equation for the function $f(x,y,t)$ in the form 
\begin{align} \label{B12c6}
f_{xx} + \frac{\gamma}{\beta} f_{yy} -f_{tt} & -\alpha
\left[ (2f_x f_{xt}+f_t f_{xx})+ \frac{\gamma}{\beta}(2f_y f_{yt}+f_t f_{yy}) \right] -\delta \left[(h f_x)_x + (h f_y)_y \right]\nonumber \\ & -\frac{1}{6} \beta\left[ f_{4x}+ \left(\frac{\gamma}{\beta}\right)^{2} f_{4y} \right] + \frac{1}{2} \beta f_{xxtt} +\frac{1}{2} \gamma f_{yytt} - \frac{1}{3} \gamma f_{xxyy} \\ &
-\alpha^{2}\left[ \frac{3}{2}\left(f_{x}^{2}f_{xx} + \left(\frac{\gamma}{\beta}\right)^{2} f_{y}^{2}f_{yy}\right) +\frac{1}{2}\frac{\gamma}{\beta}(f_{x}^{2}f_{yy} +f_{y}^{2}f_{xx})+4 f_{x}f_{y}f_{xt})  \right]
=0. \nonumber
\end{align}
Equation (\ref{B12c6}) is the single nonlinear wave equation for auxiliary function $f$ determining the velocity potential, second order in small parameters for the case $~\beta=O(\alpha^2), ~\gamma=O(\alpha^2), ~\delta=O(\alpha^2)$, when surface tension effects are negligible $(\tau\ll 1)$.

\section{Conclusions}\label{concl}

In the model of an ideal fluid, we have derived several sets of Boussinesq's equations for (2+1)-dimensional nonlinear waves. When restricted to (1+1)-dimensions, the cases discussed include the Korteweg-de Vries equation, extended KdV, fifth-order KdV, and Gardner equations, both for a flat and an uneven bottom. Even though these Boussinesq's equations are only first or second order in small parameters, they are too complicated in (2+1)-dimensions to obtain single wave equations for surface profile function $\eta(x,y,t)$. 

The Boussinesq equations, in each case, constitute a set of coupled partial differential equations for two functions, $\eta(x,y,t)$ - surface profile function and the function  $f(x,y,t)$ which determines the velocity potential through the equation (\ref{Szer1}). For (1+1)-dimensions it is always possible to eliminate $f(x,y,t)$ from the Bouusinesq set and arrive to a single wave equation of the KdV-type for the wave profile function $\eta(x,y,t)$. 
It seems to be impossible for (2+1)-dimensional cases due to the complexity of these equations. However, for each (2+1)-dimensional case, it is possible to eliminate the function $\eta(x,y,t)$ and obtain a single nonlinear wave equation for the function $f(x,y,t)$. The disadvantage of this result is that the function $f(x,y,t)$ is not a directly observed quantity. Only by knowing the $f(x,y,t)$ solution can one construct a solution of a wave profile $\eta(x,y,t)$.

Another possible approach is based on a consistent application of perturbation theory. We start with zero-order solutions. We then use the first-order Boussinesq equations to determine the first-order corrections, and if necessary, we determine further corrections from the second-order Boussinesq equations. In each of these steps, one has to solve several inhomogeneous wave equations.
This approach seems to be the most promising.

\appendix
\section{Surface tension terms} \label{appA}

In the original, dimension variables the term originating from additional pressure due to the surface tension has the following form \cite{AN2010}
\begin{equation} \label{st0}
\text{ST}=-\frac{T}{\varrho}\;\nabla \cdot \left(\frac{\nabla \eta}{(1+\vert \nabla \eta\vert^2)^{1/2}} \right) = -\frac{T}{\varrho} \frac{(1+\eta_x^2)\eta_{yy}+(1+\eta_y^2)\eta_{xx}-2\eta_x \eta_y \eta_{xy}}{(1+\eta_x^2+\eta_y^2)^{3/2}} .
\end{equation}
In (\ref{st0}), $\nabla=(\partial_x,\partial_y)$ is two-dimensional gradient operator, and  $T$ is fluid's surface tension coefficient. After transformation (\ref{bezw}) to non-dimensional scaled variables we obtain (tildas are again dropped)
\begin{align} \label{st0t}
\text{ST} & = -\frac{T}{\varrho g H^{2}}\; \frac{ \beta\left(1+\alpha^{2}\beta\eta_{y}^{2}\right)\eta_{xx}+\gamma\left(1+\alpha^{2}\beta\eta_{x}^{2}\right)\eta_{yy}-2\alpha^{2}\beta\gamma\eta_{x}\eta_{y}\eta_{xy} }{ \left(1+\alpha^{2}\beta\eta_{x}^{2}+\alpha^{2}\gamma\eta_{y}^{2}\right)^{3/2}} \nonumber \\ & 
=-\tau \left[ \beta\left(1+\alpha^{2}\beta\eta_{y}^{2}\right)\eta_{xx}+\gamma\left(1+\alpha^{2}\beta\eta_{x}^{2}\right)\eta_{yy}-2\alpha^{2}\beta\gamma\eta_{x}\eta_{y}\eta_{xy} \right] \nonumber \\ & \hspace{6ex} \times 
 \left(1 -\frac{3}{2}\alpha^{2}\beta\eta_{x}^{2}- \frac{3}{2}\alpha^{2}\gamma\eta_{y}^{2}+ O(\beta^{6}) \right)
\\ & =-\tau ( \beta\eta_{xx}+\gamma \eta_{yy} +O(\beta^4)).  \nonumber
\end{align}
In (\ref{st0t}), we first expand the factor 
$1/\left(1+\alpha^{2}\beta\eta_{x}^{2}+\alpha^{2}\gamma\eta_{y}^{2}\right)^{3/2}$
assuming that terms $\alpha^{2}\beta\eta_{x}^{2}$ and $\alpha^{2}\gamma\eta_{y}^{2}$ are small. 

Since we are interested in Boussinesq's equations up to second order in small parameters, it is enough to use the surface term in the form
\begin{equation} \label{stap}
\text{ST}  = -\tau\, ( \beta\eta_{xx}+\gamma \eta_{yy} ).
\end{equation}

\section{First step towards solving the equation (\ref{rFF})} \label{appB}

We will try to find solution to equation (\ref{rFF}) in two steps. First, assume that the function $f^{(0)}$ fulfils the equation (\ref{rFF}) reduced to zeroth order, that is, the following holds
\begin{equation} \label{kdv0}
f^{(0)}_{xx} + \frac{\gamma}{\beta} f^{(0)}_{yy} -f^{(0)}_{tt} =0.
\end{equation}
Next, postulate the solution to (\ref{rFF}) in the form
\begin{equation} \label{kdv1}
 f=f^{(0)}+\alpha\, a(x,y,t)+\beta\, b(x,y,t)+\gamma\, g(x,y,t)+\delta\, d(x,y,t),
\end{equation} 
where $a, b, g, d$ are first order correction functions. Inserting $f$ in the form (\ref{kdv1}) into (\ref{rFF}) and retaining terms up to first order in  small parameters yields
\begin{align} \label{r23a}
f^{(0)}_{xx} + \frac{\gamma}{\beta} f^{(0)}_{yy} -f^{(0)}_{tt}
& +\alpha \left( a_{xx} +\frac{\gamma}{\beta} a_{yy} -a_{tt}
   -2  f^{(0)}_{x}  f^{(0)}_{xt}  - f^{(0)}_{t}  f^{(0)}_{xx} 
   -\frac{\gamma}{\beta} \left(2f^{(0)}_{y} f^{(0)}_{yt} +f^{(0)}_{t} f^{(0)}_{yy}\right)\right) \nonumber \\ &
+ \beta  \left( b_{xx} +
   \frac{\gamma}{\beta} b_{yy} -b_{tt} +\frac{1}{2} f^{(0)}_{xxtt} 
   -\frac{1}{6}  f^{(0)}_{4x} \right) \\ & + \delta  \left(
    d_{xx} + \frac{\gamma}{\beta} d_{yy} - d_{tt}
   - \frac{\gamma}{\beta} \left( h\, f^{(0)}_{y}\right)_{y}
   -\left( h\, f^{(0)}_{x}\right)_{x} \right)  \nonumber \\ & + \gamma  \left(
   g_{xx} +\frac{\gamma}{\beta} g_{yy} - g_{tt}
   -\frac{\gamma}{\beta} f^{(0)}_{4y} 
   +\frac{1}{2} f^{(0)}_{yytt} 
   -2 f^{(0)}_{xxyy}       \right) =0.  \nonumber 
\end{align}
Since $f^{(0)}$ fulfils zeroth order equation (\ref{kdv0}) and small parameters can be arbitrary, the condition (\ref{r23a}) is equivalent to four inhomogeneous wave equations for the correction functions
\begin{align} \label{ra}
a_{xx} +\frac{\gamma}{\beta} a_{yy} -a_{tt} & =
   2  f^{(0)}_{x}  f^{(0)}_{xt}  + f^{(0)}_{t}  f^{(0)}_{xx} + 
   \frac{\gamma}{\beta} \left(2f^{(0)}_{y} f^{(0)}_{yt} 
   +f^{(0)}_{t} f^{(0)}_{yy}\right), \\ \label{rb}
b_{xx} + \frac{\gamma}{\beta} b_{yy} -b_{tt}  & = -\frac{1}{2} f^{(0)}_{xxtt} 
   +\frac{1}{6}  f^{(0)}_{4x} , \\ \label{rg}
g_{xx} +\frac{\gamma}{\beta} g_{yy} - g_{tt}   & =
   \frac{\gamma}{\beta} f^{(0)}_{4y} 
   -\frac{1}{2} f^{(0)}_{yytt} +2 f^{(0)}_{xxyy} ,  \\ \label{rd}
d_{xx} + \frac{\gamma}{\beta} d_{yy} - d_{tt}   & =
    \frac{\gamma}{\beta} \left( h\, f^{(0)}_{y}\right)_{y}
   +\left( h\, f^{(0)}_{x}\right)_{x} .  
\end{align}
For flat bottom case, $~\delta=0$, equation (\ref{rd}) does not appear.

If solutions to equations (\ref{kdv0}), (\ref{ra})-(\ref{rd}) are known,
then the time dependent surface profile is given by the equation (\ref{eFF}).


\end{document}